\documentclass{elsart3p}

\usepackage{epsfig}
\usepackage{amssymb}
\usepackage{lineno}

\RequirePackage{xspace}

\usepackage{relsize}
\def\babar{\mbox{\slshape B\kern-0.1em{\smaller A}\kern-0.1em
    B\kern-0.1em{\smaller A\kern-0.2em R}}}

\def\electron   {\ensuremath{e}\xspace}

\def\epem       {\ensuremath{e^+e^-}\xspace}
\def\muon       {\ensuremath{\mu}\xspace}
\def\g     {\ensuremath{\gamma}\xspace}

\def\qqbar {\ensuremath{q\overline q}\xspace}
\def\pion  {\ensuremath{\pi}\xspace}
\def\piz   {\ensuremath{\pi^0}\xspace}
\def\pip   {\ensuremath{\pi^+}\xspace}
\def\pim   {\ensuremath{\pi^-}\xspace}
\def\pipi  {\ensuremath{\pi^+\pi^-}\xspace}
\def\kaon  {\ensuremath{K}\xspace}
\def\KS    {\ensuremath{K^0_{\scriptscriptstyle S}}\xspace} 
\def\proton  {\ensuremath{p}\xspace}
\def\Dbar    {\kern 0.2em\overline{\kern -0.2em D}{}\xspace}

\def\Dz      {\ensuremath{D^0}\xspace}

\def\Dstarp  {\ensuremath{D^{*+}}\xspace}

\def\Bbar    {\kern 0.18em\overline{\kern -0.18em B}{}\xspace}

\def\BB      {\ensuremath{B\Bbar}\xspace} 

\mathchardef\Upsilon="7107
\def\Y#1S{\ensuremath{\Upsilon{(#1S)}}\xspace}
\def\FourS {\Y4S}

\def\pt         {\mbox{$p_T$}\xspace}

\newcommand{\gev}{\ensuremath{\mathrm{\,Ge\kern -0.1em V}}\xspace}
\newcommand{\mev}{\ensuremath{\mathrm{\,Me\kern -0.1em V}}\xspace}
\newcommand{\gevc}{\ensuremath{{\mathrm{\,Ge\kern -0.1em V\!/}c}}\xspace}
\newcommand{\mevc}{\ensuremath{{\mathrm{\,Me\kern -0.1em V\!/}c}}\xspace}
\newcommand{\gevcc}{\ensuremath{{\mathrm{\,Ge\kern -0.1em V\!/}c^2}}\xspace}
\newcommand{\mevcc}{\ensuremath{{\mathrm{\,Me\kern -0.1em V\!/}c^2}}\xspace}

\def\cm   {\ensuremath{{\rm \,cm}}\xspace}

\def\mrad {\ensuremath{\rm \,mrad}\xspace}

\def\to   {\ensuremath{\rightarrow}\xspace}

\def\pep2{PEP-II}

\newcommand{\dedx}{\ensuremath{\mathrm{d}\hspace{-0.1em}E/\mathrm{d}x}\xspace}

\newcommand{\secref}[1]{Section~\ref{sec:#1}}

\newcommand{\eqref}[1]{Eq.~(\ref{eq:#1})}

\newcommand{\DL} {\mbox{$\Delta L$}}

\usepackage{graphics}

\begin{document}

\begin{frontmatter}

\title{Extracting longitudinal shower development information from crystal calorimetry plus tracking}

\vspace{0.1cm}

\author[a]{D.N. Brown},
\ead{Dave\_Brown@lbl.gov}
\author[b]{J. Ilic},
\ead{J.Ilic@warwick.ac.uk}
\author[b]{G.B. Mohanty}
\ead{G.B.Mohanty@warwick.ac.uk}
\address[a]{Lawrence Berkeley National Laboratory, Berkeley, California 94720, USA}
\address[b]{Department of Physics, University of Warwick, Coventry CV4 7AL, United Kingdom}

\begin{abstract}
We present an approach to derive longitudinal shower
development information from the longitudinally unsegmented
\babar\ electromagnetic calorimeter by using tracking
information. Our algorithm takes advantage of the
good three-dimensional tracking resolution of \babar,
which provides an independent geometric constraint on the
shower as measured in the \babar\ crystal calorimeter.
We show that adding the derived longitudinal shower development
information to standard particle
identification algorithms significantly improves
the low-momentum separation of pions from electrons and muons.
We also verify that the energy dependence of the electromagnetic
shower development we measure is consistent with the prediction of a
standard electromagnetic shower model.
\end{abstract}


\end{frontmatter}

\section{Introduction}
\label{sec:Introduction}

In present-day nuclear and particle physics experiments,
inorganic scintillating crystals, such as NaI(Tl) and
CsI(Tl), are often used to construct electromagnetic
calorimeters when a precise measurement of the energy is
required~\cite{crystal-cal}. Crystal calorimeters can be
finely segmented in the dimension transverse to the
shower development without sacrificing energy resolution,
thus providing a good measurement of the lateral shower
development. However, engineering and energy resolution
considerations prevent finely segmenting crystal
calorimeters along the direction of shower development.
Both lateral and longitudinal
shower development information are useful in
charged particle identification (PID) algorithms,
particularly in electron identification. Because crystal
calorimeters cannot provide direct longitudinal shower
development information, they lose an important input
to particle identification.

In this paper, we present a technique in which
longitudinal shower development information is
indirectly extracted from a longitudinally unsegmented
crystal calorimeter in conjunction with a precise
tracking system. This technique
was developed for use with the \babar~\cite{babardet} detector,
but it can be applied at any detector which
combines crystal calorimetry and precision tracking.
It exploits the fact that \babar\ has a
tracking system capable of precisely determining
the three-dimensional trajectory of charged particles, and the
fact that these trajectories are not in general
collinear with the crystal axes.
A similar algorithm has previously
been used for sampling calorimeters with fine
lateral segmentation~\cite{sphagetti}.

We also explore the usefulness of the derived longitudinal shower
information for particle identification at \babar.
We show that using this
information as part of an electron identification
algorithm improves the electron {\it vs.} pion discrimination,
particularly at momenta below 600\,\mevc.
We also show that this information can be generally used to improve the
separation of the five most common charged particles
(\electron,\,\muon,\,\pion,\,\kaon\ and \proton), owing to the different longitudinal
shower development in a crystal calorimeter of these different
particle types. In particular, we find that the longitudinal shower
information significantly increases the minimum momentum reach at which
muons can be separated from pions in \babar.

We further verify that the energy
dependence of the indirect electron longitudinal shower
information behaves as expected from electromagnetic
shower models.

\section{The \babar\ Detector}
\label{sec:babar}

The \babar\ detector consists of a tracking system surrounded by
a dedicated PID device, a crystal calorimeter, and
an array of flux return iron plates instrumented with muon detectors.
A detailed description of the \babar\ detector can be found in
Ref.~\cite{babardet}, here we briefly describe those systems important
for the algorithms described in this paper.

The \babar\ tracking system is
composed of a silicon vertex tracker (SVT) comprising five
layers of double-sided detectors and a 40-layer central
drift chamber (DCH). Operating in a 1.5\,T solenoidal magnetic
field, it provides a transverse momentum ($\pt$) resolution
\begin{eqnarray}
\frac{\sigma_{\pt}}{\pt}=(0.13\pm0.01)\%\cdot{\pt}\oplus(0.45\pm0.03)\%
\end{eqnarray}
for detecting charged particles, where $\pt$ and
its rms error $\sigma_{\pt}$ are measured in \gevc.
Both the DCH and the SVT measure the specific ionization
(\dedx) of charged particles which pass through them.
The \dedx\ resolution from the DCH varies as a function of track polar
angle, having an average value of around $11\,\%$ for the majority of
physics processes that we study in \babar. The \dedx\ resolution from the
SVT is typically about $16\,\%$.

The \babar\ ring-imaging Cherenkov detector (DIRC)
provides dedicated charged particle identification
in the central part of the detector. The polar angle
coverage in the laboratory frame is $-0.84<\cos\theta<0.90$.
The Cherenkov angle resolution of the DIRC is measured to be
$2.4\,\mrad$, for the quartz refractive index of 1.473, which
provides better than $3\,\sigma$ separation between charged
kaons and pions over a broad kinematic range.

\begin{figure}[!htb]
\begin{center}
\includegraphics[width=\columnwidth]{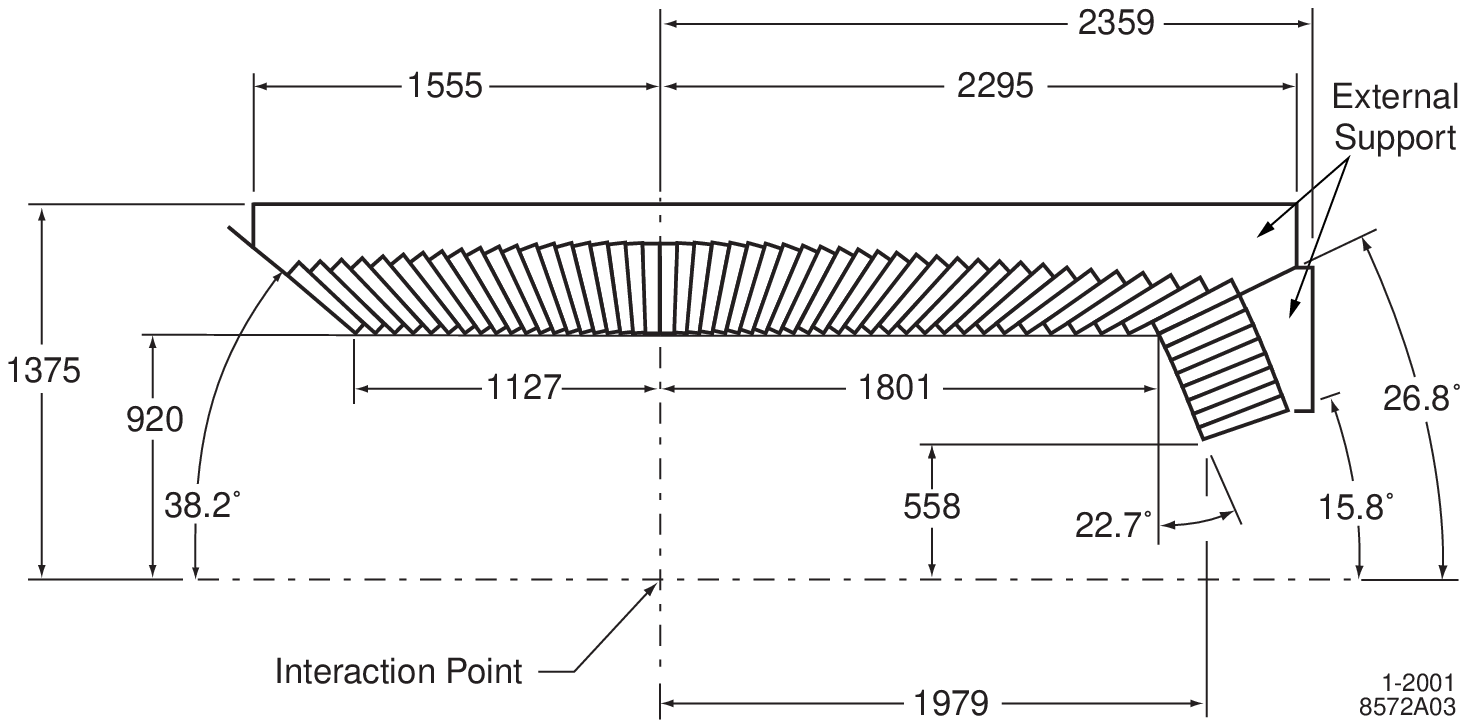}
\caption{Longitudinal cross section of the EMC showing the
top half of the detector. All dimensions are given in mm.}
\label{fig:layout}
\end{center}
\end{figure}

The \babar\ electromagnetic calorimeter (EMC) consists of
an array of 6580 CsI(Tl) crystals, which encloses the
tracking system and DIRC. The crystals have a truncated
trapezoidal shape, and are finely segmented in the plane
transverse to particles coming from the nominal
\epem\ interaction point (IP), with a typical cross-section
of $4.7\times4.7\cm^2$ at the front and $6.0\times6.0\cm^2$
at the back. The crystals range in depth
between 16 and 17.5 radiation lengths (the radiation
length of CsI(Tl) is $1.85\,\cm$), with the crystal axis
pointing back roughly to the IP. The EMC geometry can be
approximately described as a central cylindrical barrel,
divided into forward ($26.93^\circ<\theta\le90^\circ$)
and backward ($90^\circ<\theta\le140.81^\circ$) regions,
plus a conical forward endcap ($15.76^\circ<\theta\le26.81^\circ$).
The crystals are staggered so that their front face
presents a nearly normal surface to particles coming from
the IP. The EMC covers about $90\,\%$ of the polar angle
and all of the azimuth in the center-of-mass system of
the collisions produced in \pep2\ (see Fig.~\ref{fig:layout}).

The energy resolution of the calorimeter has been measured
{\em in-situ} using a combination of radioactive sources,
symmetric decays of $\piz$ and $\eta$, and Bhabha events, and
can be described as:
\begin{eqnarray}
\frac{\sigma_E}{E}=\frac{(2.30\pm0.30)\%}{\sqrt[4]{E(\mbox{GeV})}}\oplus(1.35\pm0.22)\%,
\end{eqnarray}
where $E$ and $\sigma_E$ refer to the shower energy and its
rms error, measured in \gev. The angular resolution is
limited by the transverse crystal size and the distance
from the IP. It can also be parameterized as an
energy-dependent function
\begin{eqnarray}
\sigma_\theta=\sigma_\phi=\frac{(4.16\pm0.04)}{\sqrt{E(\mbox{GeV})}}\,{\rm mrad}.
\end{eqnarray}

The EMC is surrounded by a series of iron plates arranged as coaxial
octagonal cylinders about the \babar\ symmetry axis. These plates form a
high-susceptibility path for the magnetic flux generated by the \babar\
solenoid to close on itself. Between the iron plates are resistive plate
chambers and limited streamer tubes with binary readout, used to track
muons and provide crude neutral hadron detection. The innermost layers
of muon chambers act effectively as a `tail catcher' for the EMC, detecting
particles from showers that leak out the back.

\section{Longitudinal Shower Depth Variable}
\label{sec:LongDip}

To derive longitudinal shower development information from \babar\ we
exploit the fact that most particles do not enter the calorimeter
exactly parallel to the crystal axes. A non-zero entrance angle
transforms the transverse crystal segmentation into an effective
longitudinal segmentation, providing some depth information.
Because the effective longitudinal segmentation is poor (often
fractional) and different for every particle, we do not attempt
a full parameterization of the longitudinal shower development.
Instead, we characterize the shower by the first moment of its
longitudinal development, which we call the {\em Longitudinal
Shower Depth} (\DL). The \DL\ value is closely related to,
but not identical to, the position of the electromagnetic shower
maximum, as is discussed in Appendix~\ref{sec:appendix}.

The \DL\ variable is a geometric quantity which exploits the fact that
the track and the cluster both sample different two-dimensional
projections of the three-dimensional shower spatial distribution.
When the track direction is not parallel to the crystal axis, these
projections are not fully degenerate, and they can be combined to
extract the otherwise unobservable, third (longitudinal) dimension. 

Three effects are responsible for the fact that the track direction
and the crystal axis are not collinear. First, the magnetic field
bends the track as it passes through the tracking volume. Second, the
width of the beamspot in the beam direction causes tracks from the IP to
have a different polar angle from that of the axis of the crystal they
strike. Finally, by design, the crystal axes of the \babar\ calorimeter
do not project perfectly back to the nominal IP, which reduces the chance
of particles from the IP passing perfectly between crystals.

As part of computing \DL\ we describe the calorimeter cluster as
a directed line segment in space. We first compute the two-dimensional
cluster centroid using the standard \babar\ algorithm, which
takes the weighted average of the crystal center positions
at a nominal depth of 12\,cm, using a logarithm of the crystal energy
as weight~\cite{logWeight}. We then compute the
weighted average direction of the crystal axes, using the energy in each
crystal as (linear) weight. The cluster line segment is defined to pass
through the cluster centroid, and point in the average crystal
direction. The starting point of the cluster line segment is taken as the
average position of the crystal front faces projected along the average
direction.

\begin{figure}[!htb]
\begin{center}
\includegraphics[width=.99\columnwidth]{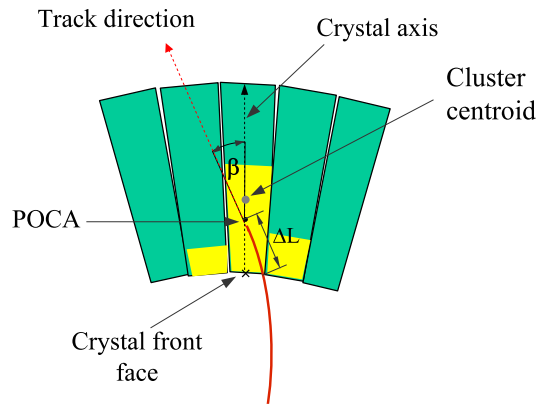}
\vskip -1.3cm
\caption{(color online) Schematic view of how \DL\ is calculated.}
\label{fig:deltaL}
\end{center}
\end{figure}

We then calculate the
point of closest approach (POCA) in three dimensions between the
extrapolated track trajectory and the cluster line segment, using an
iterative algorithm. The POCA is the point where the track
and cluster projections of the particle trajectory are most consistent.

Conceptually, we define \DL\ as the path distance the track travels
in the calorimeter's sensitive volume in reaching the POCA.
In practice, we define \DL\ as the distance along the cluster line
segment of the POCA, divided by the cosine of
the angle between the track direction and the cluster line segment
direction, given algebraically as:
\begin{eqnarray}
 \DL\equiv\frac{(\vec{r}_{\rm POCA}-\vec{r}_{\rm Front})\cdot\hat{r}_{\rm Cluster}}{\cos\beta}
\label{eqn:deltaL}
\end{eqnarray}
where $\vec{r}_{\rm POCA}$ is the position of the POCA,
$\vec{r}_{\rm Front}$ is a point on the front face of the crystal,
$\hat{r}_{\rm Cluster}$ is a unit vector in the direction of the
cluster line segment, and $\beta$ is the angle between the track
direction and the cluster axis direction.
This quantity approximates the material path distance,
but is much simpler to compute. Our definition of \DL\ ignores the
effects of track curvature and crystal-face staggering, which are
negligible on the scale of the resolution we achieve on \DL.
The definition of \DL\ is presented graphically
in Fig.~\ref{fig:deltaL}.

\section {Particle ID Control Samples}
\label{sec:samples}

We evaluate the usefulness of \DL\ for particle identification with
the control samples of relatively pure electrons, pions, kaons, muons
and protons. The considered control samples are selected from the
data collected with the \babar\ detector at the \pep2\ asymmetric energy
($3.1\gev$ on $9.0\gev$) \epem\ collider, operating near the \FourS
resonance which subsequently decays into a \BB\ meson pair. In the
following, we briefly outline the salient features of the control samples.

We select electrons\footnote{Electrons denote both electrons
and positrons.} in both radiative and non-radiative Bhabha events
[$\epem\to\epem(\g)$] by utilizing requirements on the energy
deposit and shower-shape variables in the EMC, and by rejecting
track candidates consistent with being muons. Based on Monte
Carlo studies, the purity of this sample is found to be $99.9\,\%$.
As Bhabha events provide mostly high momenta electrons, we consider
the two-photon mediated process $\epem\to(\epem)\g^*\g^*\to(\epem)\epem$
to enhance statistics in the low-to-medium momentum range ($p<3\,\gevc$).
The selection requirement for this process provides a clean sample
of electrons with purity comparable to that of the Bhabha events.

Pion candidates are selected from the decay process $\KS\to\pipi$ and
the $\epem\to\tau^+\tau^-$ events with 3-1 track topology. The purity
of the \KS\ sample selection, determined with a mixture of simulated
\BB\ decays and $\epem\to\qqbar$ continuum events, is found to be
$\sim\,99.5\,\%$. Pions from $\tau$-pair events are affected
by kaon contamination, having a purity of about
$97\,\%$. However, this sample provides high-momentum pions not found in
the $\KS\to\pipi$ sample. The electron contamination
in the $\tau$-pair pion sample is very small.

In addition to the electron and pion control samples, we use kaons
selected from the decay chain $\Dstarp\to\Dz\pip,\Dz\to K^-\pip$;
protons from the $\Lambda$ decay, $\Lambda\to p\pim$; and muons
selected in radiative muon-pair events, $\epem\to\mu^+\mu^-\g$. These
samples have the best purity for the corresponding track-candidate
selection in \babar, which is comparable to that of the considered
electron and pion samples.

\section{Electron ID Performance}
\label{sec:results}

To test the impact of \DL\ on electron identification, we start with
an electron selector based on two standard variables:
the ratio of the shower energy deposited in the calorimeter
to the momentum of the track associated with the shower ($E/p$) and
the lateral shower moment, defined as 
\begin{eqnarray}
LAT=\frac{\sum^{N}_{i=3}E_ir^2_i}{\sum^{N}_{i=3}E_ir^2_i+E_1r^2_0+E_2r^2_0}.
\end{eqnarray}
Here $N$ is the total number of crystals associated to a shower,
$E_i$ is the energy deposited in the $i$-th crystal such that
$E_1>E_2>..>E_N$, $r_i$ the lateral distance between center of the
shower and $i$-th crystal as defined earlier, and $r_0=5\,\cm$ which
is approximately the average distance between two crystals.
Like \DL, these variables involve calorimetry and tracking
measurements only.
By evaluating the incremental improvement
given by adding \DL\ to this selector, we test the impact of
\DL\ including the effect of possible correlations between
\DL\ and other similar variables.

We describe the electron identification
performance in terms of the electron efficiency and the
pion misidentification of the algorithm. We use the TMVA (Toolkit
for Parallel Multivariate Data Analysis) package~\cite{TMVA} to
build a global likelihood function using $E/p$ and $LAT$, together
(or not) with \DL.

To study the pion misidentification probability for a given value
of electron ID efficiency, first we define a likelihood ratio,
$R_L$, for each track candidate in the considered signal and
background samples by: 
\begin{eqnarray}
R_L=\frac{L_S}{L_S+L_B}.
\end{eqnarray}
Here, the signal and background likelihoods ($L_S$, $L_B$)
are products of corresponding probability density functions
($p_S$, $p_B$) of the three discriminating variables:
\begin{eqnarray}
L_{\rm S}(i) = \prod^{3}_{j=1} p_{S,j}(i).
\end{eqnarray}
After that, for a given value of the likelihood ratio, the
signal identification efficiency and the background
misidentification probability are calculated. This is done for
different momentum bins, separately in the forward and the backward
barrel, and the endcap regions,
using the control samples of electrons and pions discussed in the previous
section.

\begin{figure}[!htb]
\begin{center}
\includegraphics[width=.99\columnwidth]{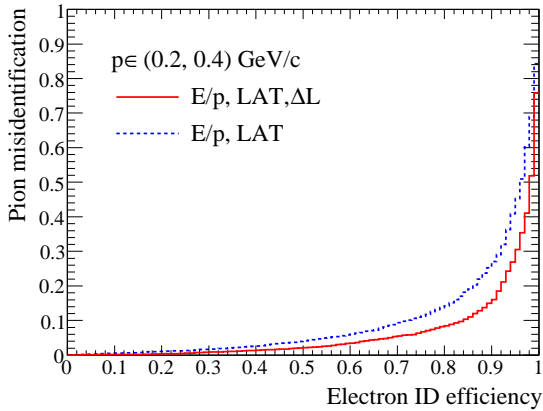}
\caption{(color online) Pion misidentification probability as a function of electron
identification efficiency in the forward Barrel region for
a typical low momentum bin: $0.2<p\le0.4\gevc$.}
\label{fig:fb}
\end{center}
\end{figure}
\begin{table}[!htb]
\caption{Comparison of pion misidentification probabilities at 90\,\%
electron identification efficiency in the case where the
likelihood function is defined with (or without) \DL.}
\label{tab:barrel_likelihood}
\center
\vskip 0.3cm
\begin{tabular}{|c|c|c|c|c|c|c|}
\hline
$p$ in &\multicolumn{2}{c|}{Backward Barrel}
 &\multicolumn{2}{c|}{Forward Barrel}
 &\multicolumn{2}{c|}{Endcap}\\
 \cline{2-7}
\gevc\      &  with  &   without  &  with  &   without  &  with  &   without  \\
\hline
 [0.2, 0.4] & 25\,\% &   34\,\%   & 16\,\% &  27\,\%    &        &            \\
\cline{1-5}
 [0.4, 0.6] & 19\,\% &   25\,\%   & 14\,\% &  22\,\%    &  5\,\% & 7\,\%      \\
\cline{1-5}
 [0.6, 0.8] &  6\,\% &   11\,\%   &  8\,\% &  15\,\%    &        &            \\
\cline{1-5}
 [0.8, 1.0] &  2\,\% &    3\,\%   &  3\,\% &   5\,\%    &        &            \\
\hline
 [1.0, 2.0] &  2\,\% &    3\,\%   &  2\,\% &   3\,\%    &  2\,\% &  3\,\%     \\
\cline{1-5}
 $>$ 2.0    &  3\,\% &    3\,\%   &  2\,\% &   2\,\%    &        &            \\
\hline
\end{tabular}
\end{table}
Figure~\ref{fig:fb} shows the electron efficiency {\it vs.} pion
misidentification probability for a typical low momentum bin ($0.2<p\le0.4\gevc$)
in the forward barrel EMC. It is evident that for any given value
of electron identification efficiency the likelihood function
based on \DL\ gives a lower pion misidentification compared to the case where
\DL\ is not included. Table~\ref{tab:barrel_likelihood}
summarizes results obtained across the full kinematic range for
various parts of the EMC. There is a clear improvement in the
performance for the backward and forward barrel regions, while
for the endcap region (where high momenta particles are mostly
abundant), we find a marginal improvement. This is because the
discrimination power of \DL\ diminishes with increasing energy.

\section{Charged Particle ID Performance}
\label{sec:pid}

\begin{figure*}[!htb]
\begin{center}
\includegraphics[width=.99\textwidth]{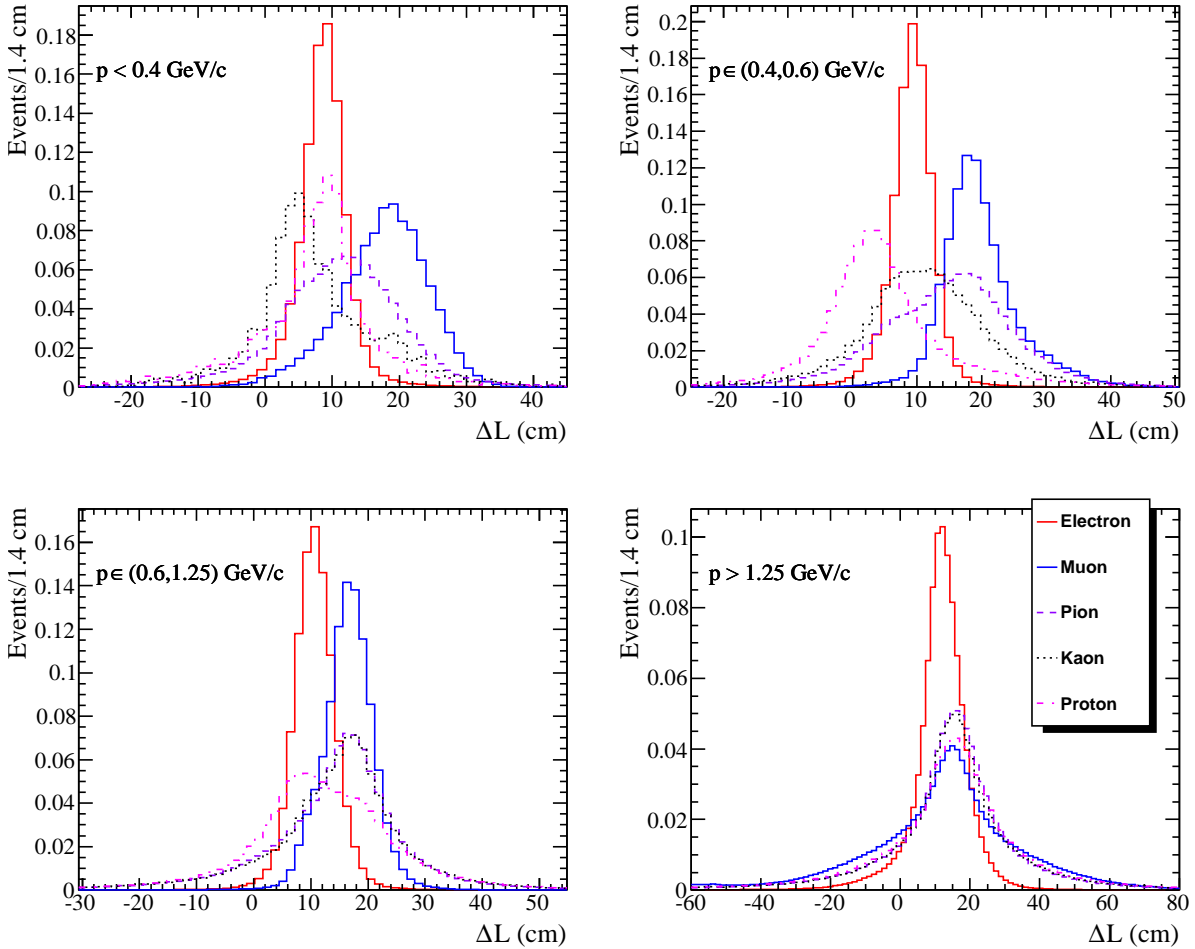}
\caption{(color online) Distributions of \DL\ for different types of
particles in different momentum bins, selected from the \babar\ data control
samples.
Note the differences in the $x$-axis range.
Each histogram has been normalized to unit area, to better show the
\DL\ distribution shapes.}
\label{fig:pidall}
\end{center}
\end{figure*}

The \DL\ variable can also be used to enhance general charged particle
identification, as it is sensitive to the differing longitudinal shower
development of different particle types. This is demonstrated
in Fig. \ref{fig:pidall}, which plots \DL\ for different species of
particles, broken down into four track-momentum bins, using the
\babar\ data PID control samples described in Section \ref{sec:samples}.

Figure \ref{fig:pidall} shows a clear distinction between the \DL\
distributions of different particle species, particularly for
momenta below 600\,\mevc.
These distributions can be basically understood in terms of the different energy
loss mechanisms at work; for instance, low momentum
protons are highly-ionizing, and so deposit most of their energy
early in the crystals. By contrast, electrons deposit their energy
near shower maximum (roughly $10\,\cm$), while
muons with momenta above 200\,\mevc are minimum ionizing and so distribute their
energy uniformly along their path through the EMC.
Finally, pions and kaons produce
broad \DL\ distributions, corresponding to the
large variability of hadronic showers.

In this momentum range \DL\ provides 0.8\,$\sigma$
pion-muon separation\footnote{We define $\sigma$ as the difference
between the average of the muon and pion \DL\ distributions
divided by the quadratic average of their rms,
$\sqrt{(\sigma^2_\mu+\sigma^2_\pi)/2}$.}, 
compared to
1.5\,$\sigma$ separation from the DIRC, less than 0.1\,$\sigma$
separation from either DCH or SVT $\dedx$~\cite{sasha},
and essentially no separation from $E/p$.
Thus \DL\ provides an useful cross-check to the DIRC when identifying
muons at these momenta, and provides the best muon-pion
separation for the 15\,\% of the \babar\ solid angle covered by
tracking and calorimetry but not by the DIRC.

At momenta above $1.25\,\gevc$, the decrease in magnetic bending
reduces the angle between the track direction and the crystal axis,
degrading the resolution of \DL. Additionally, the longitudinal profile of
energy deposition for different particle types tends to converge in
this momentum region. Some separation power still comes from
different widths of \DL\ distributions for electrons compared
to other particles, but this is a weak discriminant compared to other
PID variables available in this momentum region.

The impact of \DL\ on muon identification at \babar\ 
has been evaluated using a powerful
muon selection algorithm which uses a {\em bagged decision tree}~\cite{spr}
to combine many input variables\footnote{This muon selector has 30 input variables.}.
This algorithm was trained and evaluated
using independent subsets of the data control samples described in Section \ref{sec:samples}.
Compared to an older algorithm which does not use \DL,
the minimum muon momentum for which the selector has at least 50\,\% efficiency
(at a fixed pion misidentification probability) was reduced from 800\,\mevc to 270\,\mevc~\cite{sasha}.
This improvement in low momentum muon selection efficiency is expected to have a significant
impact on several important \babar\ physics measurements.

\section{Energy Evolution of Shower Depth}
\label{sec:depthen}

Due to the exponential nature of electromagnetic shower
development, we expect the shower maximum position
to evolve logarithmically with energy. A standard model describing
the logarithmic energy dependence of longitudinal development
is given in Ref.~\cite{PDG06}. Because
\DL\ is closely related to shower maximum, we also expect that to
evolve logarithmically with energy as well.

\begin{figure}[!htb]
\begin{center}
\includegraphics[width=0.98\columnwidth]{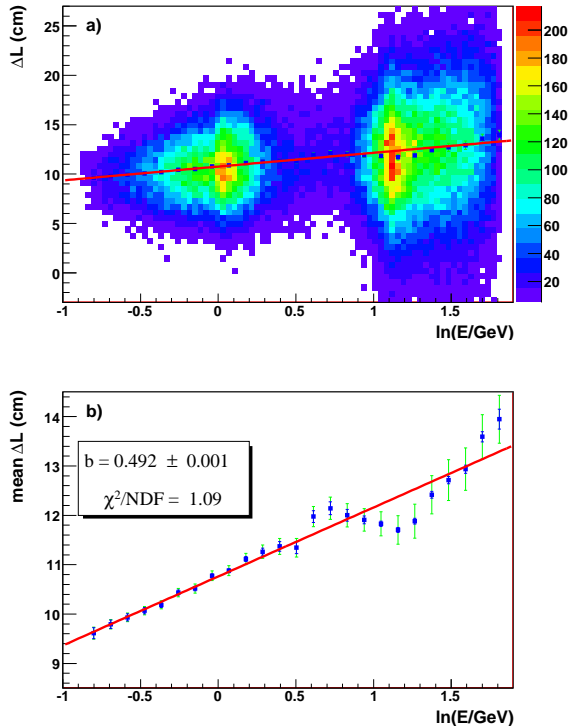}
\caption{(color online) a) Dependence of \DL\ on energy
for electrons in the \babar\ data two-photon 
and Bhabha control
samples. The points represent
the mean from a Gaussian fit to $x$-slices of the
full histogram. In b) The same points are shown with an expanded y scale.
The overlaid fit is to a shower model function described in the text.}
\label{fig:eflt}
\end{center}
\end{figure}

Figure~\ref{fig:eflt} shows \DL\ for electrons against the natural log
of the electron energy in the \babar\ data two-photon and Bhabha control
samples described
in Section~\ref{sec:samples}. The electron energy is estimated as the
track momentum calculated at the point where it enters the
EMC. The two-photon data cluster at low energy, the Bhabha at high.
The points are from slicing the data along the
horizontal axis, fitting each slice to a Gaussian, and plotting
the Gaussian mean
value with its error. The points are fit to a function based on
the shower model parameterization of \DL\ for electrons described in
Appendix \ref{sec:appendix}. This function has two free parameters:
the material-dependent shower scale factor `$b$', plus
an overall shift in \DL. The fit quality is reasonable,
showing consistency of our data with the model. The shift value we obtain
of 0.3\,cm is consistent with that expected due to
the roughly 0.2\,\% per cm light attenuation
measured in the \babar\ CsI(Tl) crystals.
The value of the shower scale factor we obtain is
$b = 0.492 \pm 0.001$, in good agreement with the value expected for CsI(Tl) of 
$0.492\pm 0.010$, where the error on the expected value comes from interpolation
in atomic number of the data provided in Ref.~\cite{pdg_emshower_bvalue},
plus the small residual energy dependence of this
parameter over the relevant energy range.

The points in Fig.~\ref{fig:eflt} show both the
statistical (small bars) and total (statistical plus systematic, larger bars) errors.
The dominant systematic error
comes from the alignment of the EMC with respect
to the \babar\ tracking system, which we estimate has an uncertainty of
$0.2$\,mrad in the rotation
around the beam axis.
Because this misalignment can depend on the
track polar angle, and because the boost of \pep2\ correlates energy with
polar angle, we include this systematic as a point-by-point error.
Other systematic effects are
from hadron backgrounds in the sample, the
uncertainty in the amount of material before the calorimeter, and
changes in light attenuation in the crystals due to radiation damage.
These latter are largely accommodated by allowing the depth offset
parameter in the fit to float.

\section{Monte Carlo Performance of \DL}
\label{sec:montecarlo}

We have also studied \DL\ in Monte Carlo
events simulated in the \babar\ detector, and compare it with real data.
This provides a useful test of the shower development model used in simulation.

\begin{figure}[!htb]
\begin{center}
\includegraphics[width=.98\columnwidth]{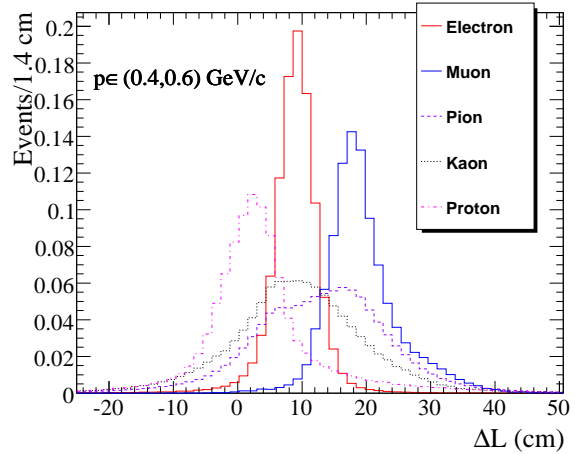}
\caption{(color online) Distributions of \DL\ for different types of
particles in the intermediate momentum bin $0.4<p<0.6\gevc$
produced in Monte
Carlo \BB\ events processed by the \babar\ detector simulation.
The histograms are normalized to unit area to better display
their shapes.}
\label{fig:pidmc}
\end{center}
\end{figure}
In Fig.~\ref{fig:pidmc} we show \DL\ distributions for different
species of particles selected in simulated control samples for the
intermediate momentum bin ($0.4<p<0.6\gevc$),
chosen because it shows maximum separation among various particles as
discussed in \secref{pid}. The Monte Carlo \DL\ shapes
are very similar to those observed in \babar\ data
(see Fig. \ref{fig:pidall}), in this and other momentum regions.

Furthermore, we have verified the logarithmic dependence of \DL\ on
the incident electron's energy in simulated samples of electrons
from Bhabha and two-photon processes.
Fitting the \DL\ {\it vs.} $\ln(E)$ distribution of these simulated electrons to
the shower model function described in 
Appendix \ref{sec:appendix}
gives a value for the shower scale factor $b = 0.493 \pm 0.001$,
consistent with expectations for a CsI(Tl) calorimeter and with our
measurements in \babar\ data.

\section{Conclusions}

In this paper, we have presented a technique for extracting longitudinal
shower development information from a crystal calorimeter in conjunction
with a precision tracking system.
When the derived quantity \DL\ is used in electron identification, we have shown that the algorithm
performance is significantly enhanced, especially in the low momentum region.
We have also shown that \DL\ can significantly improve particle separation for
other types of particles, particularly muons and pions, in the low
momentum region. By studying the energy evolution of
\DL\ for electrons in \babar, we have established that it behaves consistently
with expectations from a standard longitudinal shower model computation. We find
that the \babar\ simulation well reproduces the observed \DL\ behavior.

\section{Acknowledgments}

The work presented in this paper could not have been accomplished without the
help of many people. Many of the ideas were germinated and polished during
our inspiring discussions with Helmut Marsiske. We also thank Paul Harrison,
Chris Hawkes, Martin Kocian, Milind Purohit, and the members of the \babar\
EMC and PID subgroups for their helpful comments and suggestions throughout
this study; and \babar\ and \pep2\ for providing the data. The authors
acknowledge the support from DOE and NSF (USA), and STFC (United Kingdom).
Part of this work was supported by the Director, Office of Science,
Office of High Energy Physics, of the U.S. Department of Energy under
Contract No. DE-AC02-05CH11231.

\appendix

\section{Energy-weighted shower depth}
\label{sec:appendix}

The mean longitudinal profile of the energy deposition in an electromagnetic
shower can be well described by a gamma distribution~\cite{PDG06}:
\begin{eqnarray}
\frac{dE}{dt}=E_0\,f(t)\,\,\, {\rm with}\, f(t)=\frac{1}{\Gamma(a)}\,(bt)^{a-1}\,b\,e^{-bt},
\label{eqn:1}
\end{eqnarray}
where $t$ is the distance measured in units of radiation length $X_0$, $E_0$
is the energy of the incident particle and $E$ is the energy deposited by
the particle at a certain distance $t$. Here $b$ is a material-dependent
shower scale factor~\cite{pdg_emshower_bvalue}, and
$a \equiv (1+b\,t_{\rm max})$, where $t_{\rm max}$ denotes the shower maximum and
is expressed as follows:
\begin{eqnarray}
t_{\rm max}=\ln(E_0/E_c)+C_j,\,\,\, j = e,\gamma,
\label{eqn:2}
\end{eqnarray}
where $C_e=-0.5$ for electron-induced showers and $C_\gamma=+0.5$ for
photon showers. To a good approximation, the critical energy $E_c$ is given
as $E_c=0.8\,{\rm GeV}/(Z+1.2)=0.0145$\,GeV [for CsI(Tl), average $Z=54$].

We can then solve for the energy-weighted longitudinal shower depth, \DL, by
integrating the fractional energy deposition per radiation length in
Eq.~(\ref{eqn:1}), $f(t)$, with the corresponding path length $t$:
\begin{eqnarray}
\DL\ =\int_0^Lt\,f(t)\,dt=\frac{1}{\Gamma(a)}\int_0^L(bt)^a\,e^{-bt}\,dt.
\label{eqn:3}
\end{eqnarray}
Here $L$ indicates the average length of the CsI(Tl) crystal in units of $X_0$
and varies between 16 to 17.5 for various regions of barrel and endcap EMC.
For simplicity, we took $L$ to be 17.

We note that the \DL\ expression in Eq.~(\ref{eqn:3})
has a roughly logarithmic energy dependence [see the definition of $a$ and Eq.~(\ref{eqn:2})],
and that the shower scale factor $b$ is roughly independent of energy.
We can therefore approximate the energy and material dependence of \DL\ by
Taylor expanding Eq.~(\ref{eqn:3}) in terms of $\ln(E)$ and $b$.
Keeping only the leading linear terms:
\begin{eqnarray}
\DL \approx 10.4-4.9\Delta b+(0.9-66.0\Delta b)\ln(E),
\label{eqn:4}
\end{eqnarray}
where $\Delta b$ is the deviation of $b$ from a nominal value of
$0.5$. This expression can be used
to describe the \DL\ dependence on $\ln(E)$, for material with 
effective $Z$ values near that of CsI(Tl).

\end{document}